\documentstyle[12pt,emlines2]{article}

\title{
\begin{flushright}
{\normalsize Yaroslavl State University\\
             Preprint YARU-HE-93/02\\
             hep-ph/9404290} \\[15mm]
\end{flushright}
Electromagnetic catalysis of the radiative transitions of $\nu_i \rightarrow
\nu_j \gamma$ type in the field of an intense monochromatic wave}

\author{A.A.Gvozdev, N.V.Mikheev and L.A.Vassilevskaya\thanks{E-mail:
physteo@univ.yars.free.msk.su}\\
{\small\it Division of Theoretical Physics, Department of Physics,}\\
{\small\it Yaroslavl State University, Sovetskaya 14, 150000
Yaroslavl,}\\ {\small\it Russia}}

\date{17 September 1993}

\begin{document}

\maketitle

\begin{abstract}

The radiative decay of the massive neutrino $\nu_i \rightarrow \nu_j
\gamma$ in a circularly polarized
electromagnetic wave is investigated within the Standard theory with
lepton mixing. The decay probability in the wave field does not
contain a threshold factor $\sim ( 1 - m_i / m_j )$ as opposed to the
decay probability in a vacuum or in a constant uniform external field.
The phenomenon of the gigantic enhancement ( $\sim 10^{33}$ ) of the
neutrino decay probability in external wave field is discovered.
The probability of the photon splitting into the neutrino pair is
obtained.

\end{abstract}

\newpage

\section{Introduction}

For quite a long time there has been a stable interest in  the
physics of the massive  neutrino.  Investigations  in  this  area
touch, to a certain extent, the  feasible  phenomenon  of  lepton
mixing and  its  manifestations  in  neutrino  physics  and,
consequently,   in   astrophysics   and    cosmology    (neutrino
oscillations~[1] and their connection with the problem of the
solar neutrino~[2,3], the massive  neutrino decay~[4]
and its influence on the relict radiation spectrum~[5],
the nature of the Supernova 1987A neutrino outburst~[6],
$\beta$-decay and the problem of the 17-keV neutrino~[7].

On the other hand,  it  is  well  known  that  an  intensive
electromagnetic field can significantly influence the  properties
of the massive neutrino itself~[8]
and even induce novel lepton transitions with flavor violation,
forbidden in a vacuum~[9].
A curious, in our opinion, effect of  enhancing  influence  of  the
magnetic field on the probability of the radiative  decay
$\nu_i \rightarrow \nu_j \gamma$ ($i \neq j$, $i$~and~$j$
enumerate the definite mass  neutrino  species)  of
the massive neutrino was discovered in our recent work~[10]
in the framework of the  Glashow-Weinberg-Salam  (GWS)  theory  with
lepton mixing. We stress that lepton mixing, similarly  to  quark
mixing,  appears  quite  natural  if   the   neutrinos   have a
non-degenerate mass spectrum, and, in  itself, does not go beyond
the framework of the standard electroweak theory.

In addition the  development  of  intensive  electromagnetic
field generation techniques  and  the  current  possibility   to
obtain waves of high strength of electromagnetic field, namely
${\cal E} \sim 10^{9} V/cm$,  stimulate the investigation of quantum
processes in strong external fields. Indeed, the  parameter  of  the
wave intensity

\begin{equation}
x^2_e = - \; {e^2 a^2 \over m^2_e}
\end{equation}

\noindent (where {\it a} is the amplitude of wave, $m_e$ is the
electron mass and $e$ is  an  elementary  charge)  characterizing
the  effect of the electromagnetic wave should not be neglected.

In the present work we investigate the effect of a circularly
polarized wave on radiative decay $\nu_i \rightarrow \nu_j \gamma$
in  the  framework of the GWS theory with lepton mixing.

\section{The amplitude of the process}

In the lowest order of the  perturbation  theory,  a  matrix
element of the radiative decay of the  massive  neutrino  in
the Feynman gauge is described by diagrams of   three  types,
represented in Fig.1, where double lines imply the  influence  of
the external field. For the propagators of intermediate particles
(the $W$-boson, charged scalar and charged lepton) exact  solutions
are used of the corresponding wave equations in the  field  of  a
monochromatic circularly polarized wave with the four-potential

\begin{equation}
A_\mu = a_{1\mu} \cos\varphi  + \xi \; a_{2\mu} \sin\varphi ,
\qquad \varphi = kx
\end{equation}

\noindent where $k^\mu = ( \omega,\vec k )$ is the four-wavevector;
$k^2 = ( a_1 k ) = ( a_2 k ) = (a_ 1 a_ 2 ) = 0, \; a^2_1 = a^2_2 =
a^2$; the parameter $\xi = \pm 1$ indicates the  direction  of  the
circular polarization (left- or rightward). Note that vectors
$\vec a_1$, $\vec a_2$ and $\vec k$ form a right-handed  coordinate
system. Provided that $e{\cal E}/m^2_W \ll 1$, the main contributions
are made by the diagrams with the virtual $W$-boson in Fig.$1a$ and
the virtual $Z$-boson in  Fig.$1c$. We  stress,  that  diagram
represented in Fig.$1c$ gives the contribution to the amplitude with
$i = j$ only. This is due  to  the fact  that  flavour-changing
neutral  currents  are  absent  in  Standard Model.

The $S$-matrix element of the given process can be represented
in the following form:

\begin{equation}
S = S_0 + \Delta S
\end{equation}

\noindent where $S_0$ is the well known matrix element of the
radiative decay of  the  massive  neutrino  in  vacuum [4],
and $\Delta S$ is the contribution, induced by the wave field:

\begin{equation}
\Delta S = {i (2\pi)^4 \over \sqrt{2E_1 V \cdot 2E_2 V \cdot 2q_0 V}}
\sum^{+2}_{n=-2} {\cal M}^{(n)} \delta^{(4)}
\big ( nk + p_1 - p_2 - q \big )
\end{equation}

\noindent Here $p_1, p_2, q$ and $E_1, E_2, q_0$ are the four-momenta
and energies of the initial, final neutrinos and photon, respectively.
$n = 0, \pm 1, \pm 2$ is the difference between the numbers of absorbed
and emitted photons of the wave field.

Note that the matrix element of some process in the field of
an electromagnetic wave has usually the form of  summation  of $n$
type (4), where $- \infty < n < + \infty$ [11].
That only five values of $n$ in our case are possible is extraordinary
and is due to the following reasons. The process $\nu_i \rightarrow
\nu_j \gamma$ is local with the typical scale $\Delta x \le 1/m_f$
($m_{f}$ is the mass of the virtual fermion). In this case the angular
momentum conservation degenerates to spin conservation. Since the total
spin of the particles participating in this process is no greater
than 2 (1/2 + 1/2 + 1), $\mid n\mid_{\max} = 2$ is the maximum
difference between the numbers of absorbed and emitted photons of the
external field (the photons of a monochromatic circularly polarized
wave have a definite spin $\xi = \pm 1)$. The direct calculation
supports this conclusion. A similar phenomenon has been discovered
before [9] in studies of the effect of a circularly polarized wave
on flavor-changing transitions of the massive neutrinos
$\nu_i \rightarrow \nu_j$ ( $i \neq j$ ) with $\mid n\mid_{\max} = 1$.

Notice that in the uniform constant fields  the  decay
$\nu_i \rightarrow \nu_j \gamma$ with $m_i > m_j$
is valid only [10]. This is due to the fact that the energy-momentum
conservation law in these fields coincides with one in vacuum.
On the other hand, as it follows from expression (4), the external
electromagnetic wave field can induce also radiative decay with
$m_ i \le  m_j$  forbidden  without the field. Indeed, from the
energy-momentum conservation law in the wave field

\begin{eqnarray}
nk + p_1 = p_2 + q \nonumber
\end{eqnarray}

\noindent the relation follows

\begin{eqnarray}
m^2_i - m^2_j \ge  - \; 2n ( k p_1 ) \nonumber
\end{eqnarray}

\noindent In such a manner, the radiation decay $\nu_i \rightarrow
\nu_j \gamma$ with $m_i \le m_j$ is possible on condition that $n > 0$.

The exact invariant amplitudes ${\cal M}^{(n)}$ in the expression (4)
have cumbersome forms and will be published  elsewhere.  Here  we
present the amplitude  of  this  process  in  a  physically  more
interesting case of the ultrarelativistic neutrinos ($E_\nu \gg m_\nu$).
In this case the amplitudes ${\cal M}^{(n)}$ corresponding to $n \le 0$
are suppressed by the factor of the small neutrino mass $\nu_i$ and
the other amplitudes are significantly simplified and may be represented
as follows:

\begin{eqnarray}
{\cal M}^{(+1)} & \simeq & {G_F e^2 a^2 \over 2 \sqrt 2 \pi^2} \;
( j k ) \; {( \tilde f^{\ast} F ) \over ( k q )^2} \bigg \lbrace
\sum_\ell \Big ( K_{i \ell} K^\ast_{j \ell} - {1 \over 2} \;
\delta_{i j} \Big ) J_1 ( m_\ell ) \nonumber \\
& + & {3 \over 2} \; \delta_{i j} \sum_q \Big ( 2 T_{3q} Q^4_q \Big )
J_1 ( m_q ) \bigg \rbrace , \nonumber \\
{\cal M}^{(+2)} & \simeq & - \; { G_F \over 16 \sqrt 2 \pi^2} \;
( j F q ) \; {( f^{\ast} F ) \over ( k q )^2} \bigg \lbrace \sum_\ell
\Big ( K_{i \ell} K^{\ast}_{j \ell} + {1 \over 2} \; \delta_{i j} \;
g_\ell \Big ) J_2 ( m_\ell ) \nonumber \\
& - & {3 \over 2} \; \delta_{i j} \sum_q \Big ( Q^3_q g_q \Big )
J_2 ( m_q ) \bigg \rbrace , \\
j_\mu & = & \bar \nu_j ( p_2 ) \gamma_\mu ( 1 + \gamma_5 ) \nu_i
( p_1 ), \nonumber \\
F_{\mu \nu} & = & e ( k_\mu a_\nu - k_\nu a_\mu ) , \qquad
a_\mu = ( a_1 + i \xi \, a_2 )_\mu , \nonumber \\
f_{\mu \nu} & = & e ( q_\mu \varepsilon_\nu - q_\nu \varepsilon_\mu ),
\qquad \tilde f_{\mu \nu} = {1 \over 2} \; \varepsilon_{\mu \nu \alpha
\beta} f_{\alpha \beta} , \nonumber \\
g_{f} & = & 2 T_{3f} - 4 Q_f \sin^2\theta_W , \qquad f = \ell , q ,
\nonumber
\end{eqnarray}

\noindent where index $\ell$ indicates charged leptons
($\ell = e, \mu, \tau$)
and index $q$ indicates quark flavours ($q = u, c, t, d, s, b$),
$T_{3f}$ is the third component of the weak isospin and $Q_f$ is the
electric charge in units of the elementary charge, $\varepsilon_\mu$
is the polarization four-vector of the photon, $m_\ell$ and $m_q$ are
the masses of the virtual lepton and quark, respectively, $K_{i \ell}$
is the unitary lepton mixing matrix which can be parameterized similarly
to the quarks mixing Kabayashi-Maskawa matrix,

\begin{eqnarray}
J_1 ( m_f ) & = & \int \limits^1_0 {dy \over 1-y^2}
\int \limits^\infty_0 d\tau \; \tau^2 j_0 ( j^2_0 + j^2_1 ) \,
\exp ( -i ( \Phi ( m_f ) + \tau ) ) , \nonumber \\
J_2 ( m_f ) & = & \int \limits^1_0 dy \int \limits^\infty_0 d\tau \;
\tau ( j^2_0 + j^2_1 ) \, \exp ( -i ( \Phi ( m_f ) + 2 \tau ) ) , \\
\Phi ( m_f ) & = & { 4 \tau \over 1-y^2} \; {m^2_f \over ( k q )}
\Big [ 1 + x^2_f \Big ( 1 - j^2_0 ( \tau ) \Big ) \Big ] , \nonumber \\
x^2_f & = & - \; Q^2_f \; {e^2 a^2 \over m^2_f} , \nonumber
\end{eqnarray}

\noindent where $j_0 ( \tau ) = \sin\tau / \tau, \;
j_1 ( \tau ) = - j'_0 ( \tau )$
are spherical Bessel functions, $x^2_f$ is the parameter of the wave
intensity. It is easy to see, that the expressions for the amplitudes
(5) and (6) have no divergences and are gauge-invariant, as they are
expressed  in terms of the electromagnetic field tensor of the photon
$f_{\mu \nu}$  and the external field tensor $F_{\mu \nu}$.

Previously we have investigated the radiative decay
$\nu_i \rightarrow \nu_j \gamma$ in a homogeneous magnetic field [10].
It is known that the weak field case with large dynamical parameter
$\chi^2_f = e^2 ( p_1 F F p_1 ) / m^6_f$ corresponds to the crossed
field limit.
This circumstance may be a peculiar kind of test of the correctness
of our calculations in the wave field, since the monochromatic wave
also admits the crossed field limit ( $\omega \rightarrow 0$ with
fixed field strengths). As would be expected the amplitudes in these
both cases really coincide. In fact, the amplitude ${\cal M}^{(+1)}$
in Eqn.(5), describing the ultrarelativistic neutrino decay ($E_\nu
\gg m_\nu$) in the crossed field limit are consistent with the
corresponding expressions (8) of Ref. [10]
(${\cal M}^{(+2)} \rightarrow 0$ in this limit).

\section{The decay probability of the ultrarelati\-vis\-tic neutrino}

The decay probability $\nu_i \rightarrow \nu_j \gamma$ in the wave
field

\begin{equation}
w = \sum^{+2}_{n=-2} w^{(n)}
\end{equation}

\noindent in the ultrarelativistic limit ( $E_\nu \gg  m_\nu$ ) has
the form:

\begin{eqnarray}
E_\nu w^{(-2)} & \sim & O \bigg ( \alpha \; {G^{2}_{F} m^{10}_\nu
\over m^{4}_{e}} \; x^4_e \bigg ) , \nonumber \\
E_\nu w^{(-1)} & \sim & O \bigg ( \alpha \; {G^{2}_{F} m^{8}_\nu
\over m^{2}_{e}} \; x^2_e \bigg ) , \nonumber \\
E_\nu w^{(0)} & \sim & O \bigg ( \alpha \; G^{2}_{F} \; m^{2}_\nu \;
m^{4}_{e} \; x^{4}_{e} \bigg ) , \\
E_\nu w^{(+1)} & \simeq & {4 \alpha \over \pi } \; {G^{2}_{F} \over
\pi^{3}} \; m^{6}_{e} \; x^{6}_{e} \; \Big \vert K_{i e} K^{*}_{j e}
- {1 \over 2} \; \delta_{i j} \Big \vert^2 \nonumber \\
& \times & \int \limits^{+1}_{-1} dz {1-z \over (1+z)^{2}} \;
\big \vert J_{1} ( m_{e} ) \big \vert^2 , \nonumber \\
E_\nu w^{(+2)} & \simeq & {\alpha \over 4 \pi } \; {G^{2}_{F} \over
\pi^{3}} \; ( p_{1} k ) \; m^{4}_{e} \; x^{4}_{e} \; \Big \vert
K_{i e} K^{*}_{j e} + {1 \over 2} \; \delta_{i j} \; g_{e} \Big
\vert^2 \nonumber \\
& \times &  \int \limits^{+1}_{-1} {dz \over 1+z} \; \bigg [
{( 1 - \xi ) \over 2} + {( 1 - z )^{2} \over 4} \; {( 1 + \xi )
\over 2} \bigg ] \big \vert J_{2} ( m_{e} ) \big \vert^2 , \nonumber
\end{eqnarray}

\noindent where $z = \cos\theta$, $\theta $ is the angle between the
photon momentum $\vec q$ and the wavevector $\vec k$ in the center of
the mass frame of the $\nu_{j}$ and $\gamma $. Consequently, in the
ultrarelativistic limit $( q k ) \simeq ( p_{1} k ) ( 1 + z ) / 2$
needs to be substituted in the expression (5) and (6). Notice that
there is no singularity in the lower  limit $z \rightarrow  -1$
because  the integrals $J_{1}, J_{2}$  tend  to  zero  sufficiently
fast. Only the contribution of the virtual electron in the loop is kept
in the expressions (8). This is due to the fact that this contribution
dominates over the others under consideration

\begin{eqnarray}
E_\nu \omega  < 10^{16} (eV)^{2} , \nonumber
\end{eqnarray}

It should be pointed out that the decay probabilities (8) practically
do not depend on the mass of the neutrino. Consequently, the radiative
decay probabilities of a lighter neutrino into heavier one and
of a heavier neutrino into lighter one are equal.

It is of interest to compare the  expressions  (8)  with the well known
decay probability $\nu_i \rightarrow \nu_j \gamma$ without the field
[4]:

\begin{equation}
w_{0} \simeq  {27 \alpha \over 32 \pi } \; {G^{2}_{F} m^{5}_\nu \over
192 \pi^{3}} \; {m_\nu \over E_\nu} \;
\Big ( {m_{\tau} \over m_{W}} \Big )^4 \big \vert K_{i \tau}
K^{*}_{j \tau} \big \vert^{2} .
\end{equation}

This comparison shows that the very  small  GIM  suppression factor
$\sim ( m_{\ell } / m_{W} )^{4}$ is absent in the probability of the
radiative decay (8). Furthermore, there is no suppression caused by
the smallness of the mass of the neutrino in the case of $n = 1,2$.
To illustrate the enhancing influence  of  the  wave  field on the
decay probability $\nu_i \rightarrow \nu_j \gamma$ we present the
numerical estimation of the ratio of the probability
$\nu_i \rightarrow \nu_j \gamma$ from  the  high energy accelerator
in the wave field of laser type to  the  decay probability in vacuum:

\begin{equation}
R = {w \over w_0} \sim 10^{33} \bigg ( {1 eV \over m_\nu} \bigg )^6
\bigg ( {E_\nu \omega \over m^2_e} \bigg )^5 \Big ( 10^3 x^2_e
\Big )^2 ,
\end{equation}

\noindent where the parameter of the wave intensity (1) can be
represented in the following form:

\begin{equation}
x^2_e \simeq  10^{-3} \bigg ( {{\cal E} \over 10^9 V/cm} \bigg )^2
\bigg ( {1eV \over \omega} \bigg )^2 .
\end{equation}

Such a significant enhancement of the decay probability
$\nu_i \rightarrow \nu_j \gamma$ is, in our opinion, of great interest,
even in a relatively weak  electromagnetic field ($x^2_e \sim 10^{-3}$).
The results obtained in this work will be of use as to their application
in astrophysics and cosmology. For example, the crossed process $\gamma
\rightarrow \nu_i \tilde \nu_j$ of the photon splitting into the neutrino
pair is possible in the wave field (the amplitude of this process is
described by Eq. (5)). The probability of this process has the form:

\begin{eqnarray}
w & = & \frac{\alpha}{3 \pi} \; \frac{G_F^2}{8 \pi^3} \;
\frac{m_e^4}{q_0} \; x_e^4 \bigg \lbrace 8 m_e^2 x_e^2
\big \vert J_1 ( m_e ) \big \vert^2 \; \big \vert K_{i e} K_{j e}^* -
\frac{1}{2} \; \delta_{i j} \big \vert^2 \nonumber \\
& + & ( q k ) \big \vert J_2 ( m_e ) \big \vert^2 \;
\big \vert K_{i e} K_{j e}^* + \frac{1}{2} \; \delta_{i j}
\big \vert^2 \bigg \rbrace .
\end{eqnarray}

It can be treated as an additional mechanism of the energy loss by
stars etc.

\medskip
\medskip

The authors are grateful to A.V.Borisov and  V.Ch.Zhukovskii
for  fruitful  discussions  of  the  results  obtained.
The work was supported by the Russian Foundation of
Fundamental Research under Grant No.93-02-14414.

\newpage

\unitlength=0.80mm
\special{em:linewidth 0.4pt}
\linethickness{0.4pt}
\begin{picture}(141.50,65.50)
\put(21.00,39.05){\oval(8.00,3.00)[l]}
\put(21.00,36.55){\oval(4.00,2.00)[rt]}
\put(25.00,36.55){\oval(4.00,2.00)[lb]}
\put(25.00,34.55){\oval(4.00,2.00)[rt]}
\put(29.00,34.55){\oval(4.00,2.00)[lb]}
\put(29.00,32.55){\oval(4.00,2.00)[rt]}
\put(33.50,32.55){\oval(5.00,2.00)[b]}
\put(38.00,32.55){\oval(4.00,2.00)[lt]}
\put(38.00,34.55){\oval(4.00,2.00)[rb]}
\put(42.00,34.55){\oval(4.00,2.00)[lt]}
\put(42.00,36.55){\oval(4.00,2.00)[rb]}
\put(46.00,36.55){\oval(4.00,2.00)[lt]}
\emline{45.00}{37.55}{1}{47.00}{37.53}{2}
\emline{21.00}{40.55}{3}{47.00}{40.54}{4}
\put(48.00,39.05){\oval(6.00,3.00)[r]}
\emline{47.00}{40.55}{5}{48.00}{40.54}{6}
\emline{46.00}{37.55}{7}{49.00}{37.53}{8}
\put(21.50,39.05){\oval(7.00,1.00)[l]}
\put(22.00,37.55){\oval(4.00,2.00)[rt]}
\put(26.00,37.55){\oval(4.00,2.00)[lb]}
\put(26.00,35.55){\oval(4.00,2.00)[rt]}
\put(30.00,35.55){\oval(4.00,2.00)[lb]}
\put(30.00,33.55){\oval(4.00,2.00)[rt]}
\put(33.50,33.55){\oval(3.00,2.00)[b]}
\put(37.00,33.55){\oval(4.00,2.00)[lt]}
\put(37.00,35.55){\oval(4.00,2.00)[rb]}
\put(41.00,35.55){\oval(4.00,2.00)[lt]}
\put(41.00,37.55){\oval(4.00,2.00)[rb]}
\put(45.00,37.55){\oval(4.00,2.00)[lt]}
\put(46.50,39.05){\oval(7.00,1.00)[r]}
\emline{21.00}{39.55}{9}{47.00}{39.55}{10}
\emline{44.00}{38.55}{11}{47.00}{38.55}{12}
\emline{1.00}{39.55}{13}{17.00}{39.55}{14}
\emline{51.00}{39.55}{15}{68.00}{39.55}{16}
\put(34.00,40.32){\circle*{1.72}}
\put(34.00,41.83){\oval(2.00,2.15)[r]}
\put(34.00,43.77){\oval(2.00,1.72)[l]}
\put(34.17,45.70){\oval(1.67,2.15)[r]}
\put(34.17,47.64){\oval(1.67,1.72)[l]}
\put(34.17,49.57){\oval(1.67,2.15)[r]}
\put(34.17,51.51){\oval(1.67,1.72)[l]}
\put(34.17,53.23){\oval(1.67,1.72)[r]}
\emline{26.00}{42.05}{33}{28.00}{39.89}{34}
\emline{28.00}{39.89}{35}{26.00}{38.17}{36}
\emline{38.33}{42.05}{37}{40.33}{39.89}{38}
\emline{40.33}{39.89}{39}{38.33}{38.17}{40}
\put(38.00,52.55){\makebox(0,0)[cc]{$\gamma$}}
\put(11.00,43.55){\makebox(0,0)[cc]{$\nu_i$}}
\put(63.00,43.55){\makebox(0,0)[cc]{$\nu_j$}}
\put(25.00,45.05){\makebox(0,0)[cc]{$l^-$}}
\put(40.00,45.05){\makebox(0,0)[cc]{$l^-$}}
\put(34.00,27.05){\makebox(0,0)[cc]{\small{W} , $\varphi$ }}
\put(34.00,18.50){\makebox(0,0)[cc]{\big{(}\large{a}\big{)}}}
\emline{8.00}{38.05}{45}{11.00}{39.55}{46}
\emline{11.00}{39.55}{47}{8.00}{41.05}{48}
\emline{58.00}{38.05}{49}{61.00}{39.55}{50}
\emline{61.00}{39.55}{51}{58.00}{41.05}{52}
\put(94.33,34.28){\oval(8.00,3.00)[l]}
\put(94.33,36.78){\oval(4.00,2.00)[rb]}
\put(98.33,36.78){\oval(4.00,2.00)[lt]}
\put(98.33,38.78){\oval(4.00,2.00)[rb]}
\put(102.33,38.78){\oval(4.00,2.00)[lt]}
\put(102.33,40.78){\oval(4.00,2.00)[rb]}
\put(106.83,40.78){\oval(5.00,2.00)[t]}
\put(111.33,40.78){\oval(4.00,2.00)[lb]}
\put(111.33,38.78){\oval(4.00,2.00)[rt]}
\put(115.33,38.78){\oval(4.00,2.00)[lb]}
\put(115.33,36.78){\oval(4.00,2.00)[rt]}
\put(119.33,36.78){\oval(4.00,2.00)[lb]}
\emline{118.33}{35.78}{17}{120.33}{35.78}{18}
\emline{94.33}{32.78}{19}{120.33}{32.78}{20}
\put(121.33,34.28){\oval(6.00,3.00)[r]}
\emline{120.33}{32.78}{21}{121.33}{32.78}{22}
\emline{119.33}{35.78}{23}{122.33}{35.78}{24}
\put(94.83,34.28){\oval(7.00,1.00)[l]}
\put(95.33,35.78){\oval(4.00,2.00)[rb]}
\put(99.33,35.78){\oval(4.00,2.00)[lt]}
\put(99.33,37.78){\oval(4.00,2.00)[rb]}
\put(103.33,37.78){\oval(4.00,2.00)[lt]}
\put(103.33,39.78){\oval(4.00,2.00)[rb]}
\put(106.83,39.78){\oval(3.00,2.00)[t]}
\put(110.33,39.78){\oval(4.00,2.00)[lb]}
\put(110.33,37.78){\oval(4.00,2.00)[rt]}
\put(114.33,37.78){\oval(4.00,2.00)[lb]}
\put(114.33,35.78){\oval(4.00,2.00)[rt]}
\put(118.33,35.78){\oval(4.00,2.00)[lb]}
\put(119.83,34.28){\oval(7.00,1.00)[r]}
\emline{94.33}{33.78}{25}{120.33}{33.78}{26}
\emline{117.33}{34.78}{27}{120.33}{34.78}{28}
\emline{74.33}{33.78}{29}{90.33}{33.78}{30}
\emline{124.33}{33.78}{31}{141.33}{33.78}{32}
\put(107.00,41.39){\circle*{1.72}}
\put(107.00,43.33){\oval(2.00,2.15)[r]}
\put(107.16,45.26){\oval(1.67,1.72)[l]}
\put(107.16,47.20){\oval(1.67,2.15)[r]}
\put(107.33,49.13){\oval(1.33,1.72)[l]}
\put(107.16,51.07){\oval(1.67,2.15)[r]}
\put(107.33,53.01){\oval(2.00,1.72)[l]}
\put(107.50,54.73){\oval(1.67,1.72)[r]}
\emline{105.33}{34.51}{41}{107.66}{33.22}{42}
\emline{107.66}{33.22}{43}{105.33}{31.50}{44}
\put(111.33,52.78){\makebox(0,0)[cc]{$\gamma$}}
\put(84.33,37.28){\makebox(0,0)[cc]{$\nu_i$}}
\put(136.33,37.28){\makebox(0,0)[cc]{$\nu_j$}}
\put(117.33,43.28){\makebox(0,0)[cc]{\small{W} , $\varphi$}}
\put(95.33,43.28){\makebox(0,0)[cc]{\small{W} , $\varphi$}}
\put(107.33,26.78){\makebox(0,0)[cc]{$l^-$}}
\put(106.33,18.28){\makebox(0,0)[cc]{\big{(}\large{b}\big{)}}}
\emline{81.33}{32.28}{53}{84.33}{33.78}{54}
\emline{84.33}{33.78}{55}{81.33}{35.28}{56}
\emline{131.33}{32.28}{57}{134.33}{33.78}{58}
\emline{134.33}{33.78}{59}{131.33}{35.28}{60}
\end{picture}


\unitlength=0.80mm
\special{em:linewidth 0.4pt}
\linethickness{0.4pt}
\hspace{20mm}
\begin{picture}(65.00,69.75)
\emline{25.00}{30.00}{1}{65.00}{30.00}{2}
\emline{32.00}{28.50}{3}{35.00}{30.00}{4}
\emline{35.00}{30.00}{5}{32.00}{31.50}{6}
\emline{55.00}{28.50}{7}{58.00}{30.00}{8}
\emline{58.00}{30.00}{9}{55.00}{31.50}{10}
\put(45.00,31.50){\oval(3.00,3.00)[r]}
\put(45.00,34.50){\oval(3.00,3.00)[l]}
\put(45.00,37.50){\oval(3.00,3.00)[r]}
\put(45.00,40.50){\oval(3.00,3.00)[l]}
\put(45.00,50.00){\circle{16.00}}
\put(45.00,50.00){\circle{13.00}}
\emline{35.50}{48.00}{11}{38.00}{51.00}{12}
\emline{38.00}{51.00}{13}{40.50}{48.00}{14}
\emline{49.50}{51.00}{15}{52.00}{48.00}{16}
\emline{52.00}{48.00}{17}{54.50}{51.00}{18}
\put(45.00,43.00){\circle*{2.50}}
\put(50.00,64.00){\makebox(0,0)[cc]{$\gamma$}}
\put(59.00,49.00){\makebox(0,0)[cc]{\cal f}}
\put(31.00,49.00){\makebox(0,0)[cc]{\cal f}}
\put(50.00,37.00){\makebox(0,0)[cc]{Z}}
\put(33.00,25.00){\makebox(0,0)[cc]{$\nu_i$}}
\put(57.00,25.00){\makebox(0,0)[cc]{$\nu_i$}}
\put(45.00,15.00){\makebox(0,0)[cc]{\big{(}\large{c}\big{)}}}
\put(45.00,0.00){\makebox(0,0)[cc]{\large{Fig. 1.}}}
\put(45.00,58.50){\oval(3.00,3.00)[r]}
\put(45.00,61.50){\oval(3.00,3.00)[l]}
\put(45.00,64.50){\oval(3.00,3.00)[r]}
\put(45.00,67.50){\oval(3.00,3.00)[l]}
\put(45.00,57.00){\circle*{2.50}}
\end{picture}


\newpage

\begin{thebibliography}{11}

\bibitem{BU}
  J.N.Bahcall and R.K.Ulrich,
  Rev. Mod. Phys. 60 (1989) 297.

\bibitem{MS/BB}
  S.P.Mikheyev and A.Yu.Smirnov,
  Usp. Phys. Nauk 153 (1987) 3; \\
  J.N.Bahcall and H.A.Bethe ,
  Phys. Rev. Lett. 65 (1990) 2233.

\bibitem{BNKL}
  S.A.Bludman, N.Nata, D.C.Kennedy and P.G.Langacker,
  Phys. Rev. D 47 (1993) 2220.

\bibitem{LSh}
  B.W.Lee and R.E.Shrock,
  Phys. Rev. D 16 (1977) 1444.

\bibitem{AV}
  T.M.Aliyev and M.I.Vysotskii,
  Usp. Phys. Nauk 135 (1981) 709.

\bibitem{CL}
  D.O.Caldwell and P.G.Langacker,
  Phys. Rev. D 44 (1991) 824.

\bibitem{GNP}
  G.Gelmini, S.Nissinov and R.D.Peccei,
  Int. J. Mod. Phys. A 7 (1992) 3141.

\bibitem{ZhMB}
  V.Ch.Zhukovskii, I.B.Morozov and A.V.Borisov,
  Yad. Phys. 37 (1983) 698.

\bibitem{BTV}
  A.V.Borisov, I.M.Ternov and L.A.Vassilevskaya,
  Phys. Lett. B 273 (1991) 163.

\bibitem{GMV}
  A.A.Gvozdev, N.V.Mikheev and L.A.Vassilevskaya,
  Phys. Lett. B 289 (1992) 103.

\bibitem{BLP}
  V.B.Berestetskii, E.M.Lifshitz and L.P.Pitaevskii,
  Quantum electrodynamics (2nd ed., Pergamon Press, Oxford, 1982).

\end{thebibliography}
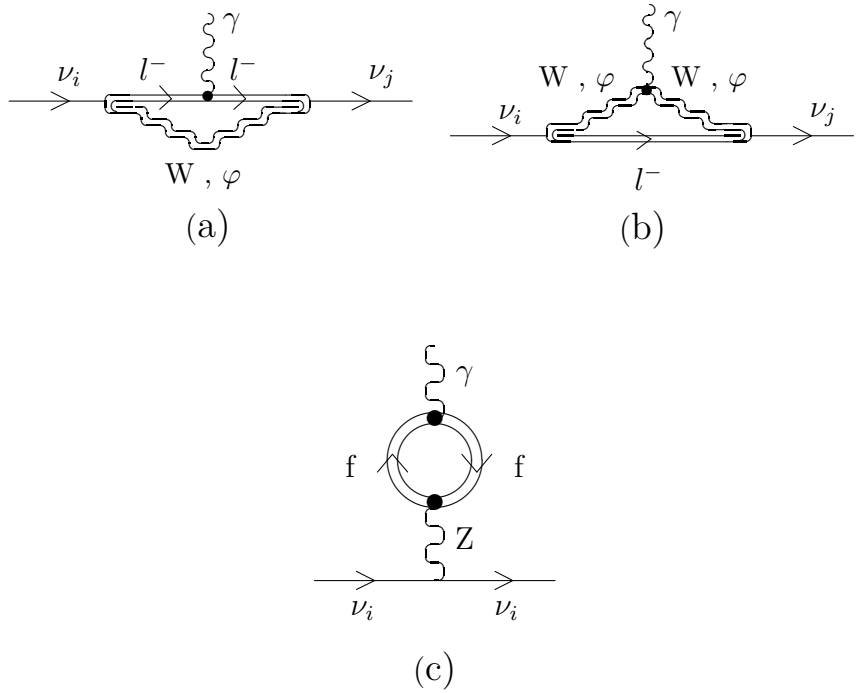
\end{document}